\documentclass[aps,pra,twocolumn,showpacs,superscriptaddress,groupedaddress]{revtex4-1}
\usepackage[margin={2cm,2cm}]{geometry}
\usepackage{graphicx}
\usepackage{amsthm}
\usepackage[english]{babel}

\usepackage{amsmath}
\usepackage{amsfonts}
\usepackage{amssymb}
\usepackage{subfig}
\usepackage{color}
\usepackage{bbold}
\usepackage{bbm}
\usepackage{amsfonts}
\usepackage{hyperref}
\usepackage{epstopdf}
\usepackage{lipsum}

\newcommand{\ket}[1]{\ensuremath{\left|#1\right\rangle}}

\begin{document}
  \title{Is non-locality stronger in higher dimensions?}
  \author{Soumik Adhikary}
  \email{phz148122@physics.iitd.ac.in}
  \affiliation{Department of Physics, Indian Institute of Technology Delhi, New Delhi-110016, India.}

  \author{V. Ravishankar}
  \email{vravi@physics.iitd.ac.in}
  \affiliation{Department of Physics, Indian Institute of Technology Delhi, New Delhi-110016, India.}
  
    \author{Radha Pyari Sandhir}
  \email{radha.pyari@gmail.com}
  \affiliation{Department of Physics and Computer Science, Dayalbagh Educational Institute, Dayalbagh Rd., Agra, Uttar Pradesh-282005, India.}
  
  \begin{abstract}
  A critique of a prescription of non-locality \cite{Kasz00,Coll02,FuLi04}  that appears to be stronger and more general than the Bell-CHSH formulation is presented.    It is shown that, contrary to expectations,  this prescription fails to correctly identify a large family of maximally non-local Bell states.

  \end{abstract}
  

\maketitle  

\section{Introduction}
\label{sec:introduction}

Violation of the Bell inequality is an unmistakable signature of non-locality for any quantum system. This has been tested extensively and spectacularly in two qubit systems. Yet the formulation, though general enough to be applied to any bipartite quantum system, throws open the question of  whether there could be non-local states which do {\it not} violate the Bell inequality. A definitive answer to this, in the negative,   is available only for two qubit systems \cite{Fine82}.The question remains to be settled for $N \times N$  systems.

In an attempt to resolve this issue, two new formulations of non-locality have been proposed by Kaszlikowski et al \cite{Kasz00} and
Collins et al \cite{Coll02} (see \cite{FuLi04} for a general formulation, \cite{Durt01,Chen01,Kasz02} for discussions of special cases). The formulations are equivalent and involve new inequalities which, of course, reduce to the standard Bell-CHSH form for two qubits. Remarkably, the inequalities -- unlike in the Bell case -- are dimension dependent, and are not constrained by the Cirel'son bound \cite{Cirelson80}. When applied to the fully entangled states and their contamination by white noise, shown below,
\begin{equation}\label{eq:wer}
\vert \Psi_E\rangle = \frac{1}{\sqrt{N}} \sum_j\vert jj\rangle; \quad \rho = p\vert \Psi_E \rangle \langle \Psi_E\vert + (1-p) \frac{\mathbb{I}}{N^2},
\end{equation}
one is led to conclude that non-locality gets stronger with increasing dimensions, and that one can identify non-local states that evade the Bell analysis. The new inequalities are also verified experimentally \cite{Howe02,Vaziri02,Thew04,Dada11,Lo16}, the last one involving coupled systems upto $N=16$. 

Thus the new inequalities hold the promise of generalising and subsuming the standard Bell analysis for identifying 
and characterising non-local states in higher dimensions. It is, therefore, an opportune moment to examine, in a greater detail,  the domain of applicability of the new inequalities. For, as mentioned, the new inequalities  and their experimental tests have been applied only on the states described by Eq. \ref{eq:wer}.  
This paper undertakes the task, for the specific case of two coupled 4-level systems, and examines whether  the new inequalities are more discriminating of non-local states vis-a-vis the Bell-CHSH, or simply Bell, inequality. In this exercise, we consider the inequality proposed in \cite{Coll02}, which we shall call as CGLMP, after the authors, and apply them on Bell states, i.e., those that violate Bell inequality maximally.

\section{Formulation}
\label{sec:formulation}
\subsection{Bell and CGLMP inequalities}
For the sake of completeness, we describe the CGLMP inequality briefly, after mentioning the standard Bell inequality.  Consider an $M \times N$ level system with two subsystems $A(B)$ of $M(N)$ levels.
 The Bell operator  $\mathcal{B}$ is defined  by,
\begin{equation} \label{eq:bell_function}
\mathcal{B} = A_1 B_1 - A_1 B_2 + A_2 B_1 + A_2 B_2
\end{equation}
where $A_{1,2}$ and $B_{1,2}$ are  local observables for the two subsystems,   subject to the conditions  $-1 \leq \big< A_i \big> \leq 1$ and $-1 \leq \big< B_i \big> \leq 1$. Local hidden variables models  constrain the Bell function to obey the inequality 
$
\big| \big< \mathcal{B} \big> \big| \leq 2,
$
a violation of which  implies non-locality. As a non-local theory, quantum mechanics pushes the upper bound to a value $2\sqrt{2}$ \cite{Cirelson80}. This bound is absolute and independent of $M$ and $N$.

 CGLMP inequality has a more complicated structure. The analog of the Bell operator is the function  $I_N$, defined for an $N \times N$ level system by 

\begin{widetext} 
\begin{align}\label{eq:cglmp_inequality}
I_N = \sum_{k=0}^{[N/2]-1} \Big(1- \frac{2k}{N-1}\Big) \{ & [P(A_1 = B_1 + k) + P(B_1 = A_2 +k+1) + P(A_2 = B_2 +k) + P(B_2 = A_1 +k)] \nonumber \\ & - [P(A_1 = B_1 -k -1) +P(B_1 = A_2 -k) +P(A_2 = B_2 -k -1) + P(B_2 = A_1 -k -1) ]\}
\end{align}
\end{widetext}
where $P(A_i,B_i)$ are joint measurement probabilities for local observables $A_i,B_i$ belonging to the two subsystems. All the observables have integer eigenvalues $0,1, \cdots, N-1$. 
The measurement prescription can be found in \cite{Durt01}, and involves two local observers,  Alice and Bob,  who fine tune variable phases $\alpha_i,\beta_i$ (see Eq. \ref{eq:basis}) of the states in their respective subsystems, depending on the measurements they wish to perform. The measurement bases for the observables $A_i$ and $B_i$; $i = 1,2$ are of  the form
\begin{eqnarray} \label{eq:basis}
\ket{K}_{A,i} = \frac{1}{\sqrt{N}} \sum_{j=0}^{N-1}{exp \Big(i\frac{2 \pi}{N} j (K + \alpha_i) \Big) \ket{j}_A} \nonumber \\
\ket{L}_{B,i} = \frac{1}{\sqrt{N}} \sum_{j=0}^{N-1}{exp \Big(i\frac{2 \pi}{N} j (-L + \beta_i) \Big) \ket{j}_B}.
\end{eqnarray}
 Rules of classical probability  impose the constraint  $\vert I_N \vert \le 2$. This constraint is interpreted as a locality condition inasmuch as it arises in measurements involving joint probabilities. 
 
 On evaluating $I_N$ for the maximally entangled state, it is found that the maximum value of $I_N$ as allowed by quantum theory transcends the Cirel'son bound when $N \ge 3$. It takes a value 2.8962 when $N=4$, and approaches the limiting value of 2.9696 as $N \rightarrow \infty$. Since $I_4 > 2\sqrt{2}$ for the Bell state, it follows that some  noisy states (defined in Eq.~\ref{eq:wer}), will obey Bell inequality but violate CGLMP, buttressing the claim that CGLMP prescription is more general  and stronger than the Bell prescription. As remarked, experiments are also performed on maximally entangled states.
 
 \section{Critique of CGLMP inequality}
We wish to contrast the CGLMP inequality with the Bell formulation for a broader class of states for a $4 \times 4$ level system. In this, we freely exploit  the abundance of Bell states which are not restricted to be  fully entangled, or even pure. We follow the analysis in \cite{Brau92} closely in the construction of Bell states.

\subsection{Bell states  of $4 \times 4$ level systems}\label{sect:example}
It is known that the conditions on the local observables for attaining the Cirel'son bound are given by \cite{Pope92}
\begin{eqnarray} \label{eq:maximal_condition}
\big< A_{1,2}^2 \big> &= & \big< B_{1,2}^2 \big> = 1 \nonumber \\
\big< \{A_1,A_2\} \big> & \rm{or} &  \ \big< \{ B_1,B_2 \} \big> = 0.
\end{eqnarray}
The two conditions jointly constitute the definition of Clifford Algebra,  the representations  of which are essentially given by the standard Pauli matrices or their direct sums for each pair of observables.  It follows thereof that maximally non-local Bell states are either coherent, or incoherent superpositions of Bell states in mutually orthogonal $2 \times 2$ sectors. This is a consequence of Jordan's theorem \cite{Pope92,Brau92}. This explains why there are no fully entangled Bell states when $N$ is odd.

 Armed with this result  we conveniently choose the observables to be 
\begin{eqnarray} \label{eq:observables}
A_1 &= & \frac{2}{\sqrt{3}} \lambda_8 + \frac{\sqrt{6}}{3} \lambda_{15}; \ \ A_2 = (\lambda_4 + \lambda_{11}) \nonumber \\
B_1 &= & \frac{1}{\sqrt{2}} (A_1 + A_2); \ \ B_2 = \frac{1}{\sqrt{2}} (A_2 - A_1)
\end{eqnarray}
where the $SU(4)$ generators (the $\lambda$ matrices) are taken in their standard form. The spectrum of the Bell operator is easily determined. 
It has the eigen-resolution 
\begin{equation}
\mathcal{B} = 2\sqrt{2}(\Pi_{\mathcal{H}_+} - \Pi_{\mathcal{H}_-})
\end{equation}
where $dim (\mathcal{H}_{\pm}) =4$. The bases for $\mathcal{H}_{\pm}$ may be chosen to be
 \begin{eqnarray} \label{eq:states_maximal_p}
\ket{\eta_1}& = &\frac{1}{\sqrt{2}} ( \ket{11} + \ket{33}) ; \ \ \ket{\eta_2} = \frac{1}{\sqrt{2}} ( \ket{10} + \ket{32}) \nonumber \\
\ket{\eta_3} &= &\frac{1}{\sqrt{2}} ( \ket{01} + \ket{23}); \ \ \ket{\eta_4} = \frac{1}{\sqrt{2}} ( \ket{00} + \ket{22}) 
\end{eqnarray}
and 
\begin{eqnarray} \label{eq:states_maximal_m}
\ket{\phi_1} &= &\frac{1}{\sqrt{2}} ( \ket{31} - \ket{13}) ; \ \ \ket{\phi_2} = \frac{1}{\sqrt{2}} ( \ket{30} - \ket{12}) \nonumber \\
\ket{\phi_3} &= & \frac{1}{\sqrt{2}} ( \ket{21} - \ket{03}) ; \ \ \ket{\phi_4} = \frac{1}{\sqrt{2}} ( \ket{20} - \ket{02})
\end{eqnarray}
respectively. Within each sector, all states, both pure and mixed, violate the Bell inequality maximally. Thus, in contrast to the two qubit case, Bell states can have ranks ranging from $1-4$. 
We consider states belonging to $\mathcal{H}_{+}$  henceforth. We examine the pure and mixed states separately in the next section.

\subsection{Comparison of Bell and CGLMP measures}
\subsubsection{Pure states}
The comparison requires a numerical study of the behaviour of $I_4$. It is convenient to represent a pure Bell state
in the form
\begin{equation}
\ket{\Psi}_{\mathcal{H}_+} = \sum_i{c_i \ket{\eta_i}} \ \ ; \ \ \sum_i {|c_i|^2} =1
\end{equation}
with the parametrisation
\begin{eqnarray} \label{eq:parameters}
c_1 &=& \cos{\theta_1}  \nonumber \\
c_2 &=& \exp(i\gamma_1)\sin{\theta_1} \cos{\theta_2} \nonumber \\
c_3 &=& \exp(i\gamma_2)\sin{\theta_1} \sin{\theta_2} \cos{\theta_3} \nonumber \\
c_4 &=& \exp(i\gamma_3)\sin{\theta_1} \sin{\theta_2} \sin{\theta_3}.
\end{eqnarray} 
The states constitute a six dimensional manifold $\mathcal{M}^6 \equiv \mathcal{S}^3 \times (\mathcal{S}^1)^{\otimes 3}$. The task consists of optimising the tunable phases $\alpha_i,\beta_i$ (Eq.~\ref{eq:basis}) in order to maximise $I_4$. For instance, numerical simulations performed in \cite{Durt01} determine the optimal values for the maximally entangled state $\ket{\Psi_E}$  to be 
$ (\alpha_1, \alpha_2)=(0,1/2); ( \beta_1,\beta_2) =(1/4,-1/4)$.This configuration yields a value of  $I_4 = 2.8962$, which is in excess of $2\sqrt{2}$.

\subsubsection*{The method}
In order find the maximum $I_4$ value for each state, we implement the Nelder-Mead optimization technique \cite{Neld65}, to search over the 4-dimensional phase parameter space spanned by  $\{\alpha_{1,2},\beta_{1,2}\}$. The same technique was used in \cite{Kasz00,Durt01} to optimise the parameters for the fully entangled state. This method does not always guarantee convergence; the search stops when the `standard error' falls under a certain pre-defined value: $\sqrt{\frac{1}{n+1}\sum^{n}_{i=0}(f(x_1)-\overline{f(x_i)})^2}$ . Its success depends on the simplex not becoming too small in relation to the curvature of the surface.  If the simplex is too small, chances of it being trapped in a local maxima are high.   However, through selection of different initial test points across the parameter space to commence the search, one can avoid repeatedly falling into the same local maxima, and a global maxima can be obtained.

\begin{figure}[!ht]
\includegraphics[width=\linewidth]{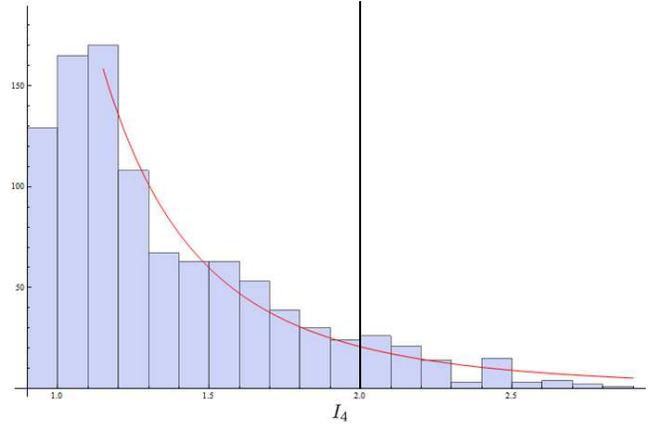}
\caption{\label{fig:histogram_pure} (colour online) $I_{4}$ values for 1000 randomly sampled pure states, with a bin-width of 0.1. Red: Polynomial fit showing population decay. Black:  $I_4=2$, demarcation between local and non-local states.}
\end{figure}

For each state in our simulations, a number of such searches were  performed.  Henceforth, we simply denote $I_4$ as the maximum value pertaining to each state across these multiple searches.  1000 pure states were sampled randomly from $\mathcal{M}^6$ through uniform distributions. The pre-defined error criterion was taken to be 0.0001.  The results are given in the histogram in Fig.~\ref{fig:histogram_pure}.

We see that most of the Bell states respect the CGLMP inequality. Out of the 1000 states, it is found that only 8.9\% violate the CGLMP inequality. A polynomial fit shows the population of the sampled states decreasing as a function of $I_4$,  at a rate $(I_4)^{-x}$, where $x = 3.66 \pm 0.49$ (see Fig.~\ref{fig:histogram_pure}). This reflects the sparsity of states that obey the CGLMP criterion for non-locality in contrast to the Bell criterion. In fact, there is only one state with $I_4 > 2\sqrt{2}$ from among the 1000 random states.

\subsubsection{Mixed states}
\begin{figure}[!ht]
\includegraphics[width=\linewidth]{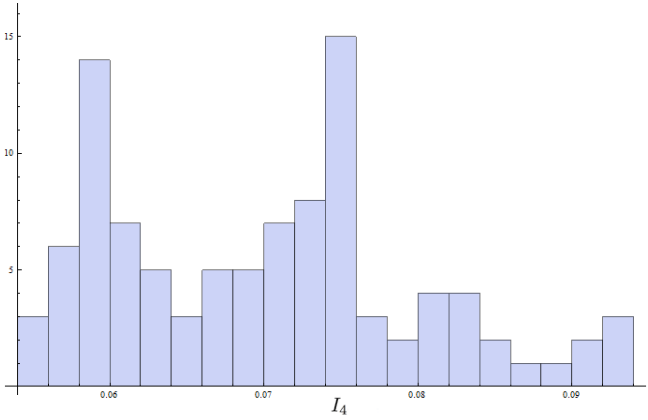}
\caption{\label{fig:Histogram_mixed} (colour online) $I_{4}$ values for 100 randomly sampled mixed states, with a bin width 0.002.}
\end{figure}

 The space of mixed states is much larger, being of dimension 15. We examine the rather small subset of states which have the basis states $\vert \eta_i\rangle$ (Eq.~\ref{eq:states_maximal_p}) as their eigenstates.
\begin{equation}
\rho_{\mathcal{H}_+} = \sum_i p_i |\eta_i \rangle \langle \eta_i|. \label{eq:mixed}
\end{equation}
Once again, we find  that the CGLMP prescription fails to identify non-locality, this time more dramatically. We sample 100 random mixed states, and implement the Nelder-Mead optimisation technique, in the same manner that was done for pure states. Fig.~\ref{fig:Histogram_mixed} shows the maximum $I_4$ values obtained for each state.    All of these states have values of $I_4$ very close to zero,  despite being maximally Bell non-local.

\begin{figure}[!ht]
\includegraphics[width=\linewidth]{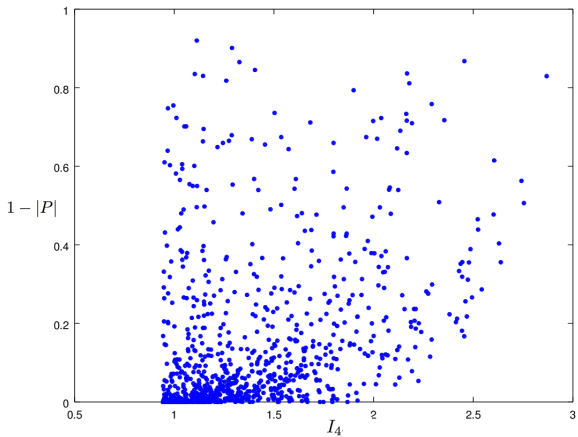}
\caption{\label{fig:ent_cglmp} (colour online) $I_{4}$ plotted as a function of entanglement, $(1- |P|)$.}
\end{figure}

It is clear from Figs. \ref{fig:histogram_pure}, \ref{fig:Histogram_mixed} that CGLMP  is not necessarily stronger than Bell, and that it fails to identify a large family of non-local states, unless perhaps they are maximally entangled.  Since it fares well for fully entangled states, we examine if $I_4$ is more sensitive to  entanglement of the state. 

\subsubsection{ CGLMP measure and entanglement}
We consider pure Bell states. The eigenvalues of the  reduced density matrix  are two fold degenerate and can be written as
\begin{equation}
 \mu_{\pm} = \frac{1}{4} (1 \pm P); ~~ |P| \le 1.
\end{equation}
The quantity $1-|P|$ is a measure of entanglement, with $|P|=0$ representing a fully entangled state, and $|P|=1$,
a partially entangled state with the corresponding entropy of the reduced density matrix  being  $\log 2$.

Fig.~\ref{fig:ent_cglmp} shows the variation of  $I_4$ with respect to $1-|P|$   for  the 1000 states employed earlier.  
The scatter in the plot clearly shows that $I_4$ bears no affinity to entanglement either,  except in the limiting case of maximum entanglement.

\section{Discussion and conclusion} 
The main result of this paper is that CGLMP inequality (and its equivalent formulation \cite{Kasz00}), as a measure of non-locality, fails to identify a large class of  Bell states, both in pure and mixed sectors. It is also not sensitive to entanglement, except when the state is also fully entangled. There are many measures of non-classicality such as discord, entanglement and non-locality. Though undoubtedly a measure of non-classicality, it is not clear whether a CGLMP violation reflects some or all of these measures, or if it gives a new measure. Strictly speaking, Bell  and CGLMP criteria are based on two independent notions of non-locality, even if they agree in the case of $2 \times 2$ systems.  Therefore, it must be admitted that the precise non-classical nature of even  those states which respect Bell but violate CGLMP still remains to be understood.
Notwithstanding these reservations, there is no doubt that the ingenious experiments which have verified CGLMP violation \cite{Howe02,Vaziri02,Thew04,Dada11,Lo16} have indeed detected a non-trivial non-classical feature
of quantum states in higher dimensions which is not available for two qubit states.

\section*{Acknowledgements}

Radha thanks the Department of Science and Technology (DST), India for funding her research under the WOS-A Women's Scientist Scheme. Soumik thanks Council of Scientific and Industrial Research (CSIR), India for funding his research.

\end{document}